# MRI of the lung using hyperpolarized [3]He at very low magnetic field (3 mT)

C.P. Bidinosti(✉)[1], J. Choukeife[2], G. Tastevin[2], A. Vignaud[3], and P.-J. Nacher[2]

1.  *Department of Physics, Simon Fraser University, Burnaby, Canada V5A 1S6*

2.  *Laboratoire Kastler Brossel, 24 rue Lhomond, F75231 Paris, France*

3.  *U2R2M, Université Paris Sud, Bat. 220, 91405 Orsay Cedex, France*

Email: cpbidino@sfu.ca

Fax: 604-291-3592

Optical pumping of [3]He produces large (hyper) nuclear-spin polarizations independent of the magnetic resonance imaging (MRI) field strength. This allows lung MRI to be performed at reduced fields with many associated benefits, such as lower tissue susceptibility gradients and decreased power absorption rates. Here we present results of 2D imaging as well as accurate 1D gas diffusion mapping of the human lung using [3]He at very low field (3 mT). Furthermore, measurements of transverse relaxation in zero applied gradient are shown to accurately track pulmonary $O_2$ partial pressure, opening the way for novel imaging sequences.

## Introduction

Magnetic resonance imaging of the lung using hyperpolarized (HP) noble gases ([3]He or [129]Xe) has advanced significantly: measurable quantities such as gas distribution, diffusion and nuclear-spin relaxation have been shown to correlate with pulmonary physiology, and clinical applications are being actively explored [1,2]. To date, almost all in vivo studies have been performed in *high field* ($B_o$ ~1.5 T) imagers, which are the most prevalent owing to the demands of conventional MRI where signal-to-noise (SNR) is proportional to $B_o$. With HP MRI, however, optical pumping techniques produce a nuclear magnetization that is independent of $B_o$ and there exists no SNR benefit to operation at 1.5 T [1]. On the contrary, HP MRI might be better served by operation at lower magnetic fields where certain benefits may exist. Most notably, local field gradients, which are due to spatial variations in tissue susceptibility, decrease in going to a lower imaging field.  Also, working at a reduced frequency reduces the RF power absorbed by the body, thereby allowing the use of rapid pulse sequences without exceeding safety limits.  Other proposed benefits have been described elsewhere [3,4], and include the development of light weight, MR imagers for orientational lung study [5].

We began exploration of very low field HP MR in the human lung using a homebuilt imager operating at 3mT [3].  Poor homogeneity of the $B_0$ field, however, limited the relevance of local diffusion measurements and prevented gas density imaging.   In this work we employed a commercial whole-body imager (Sopha-Magnetech, nominally designed for 0.1 T [6]), operating at reduced current to perform $^3$He MRI of the lung at very low field.

## Method

Measurements were made at 104.5 kHz  (3.222 mT) inside a copper mesh Faraday cage, which  attenuated external noise signals by ~ 40 dB.  A standard laboratory power supply was used to provide a stable current to the resistive magnet, and bipolar current amplifiers (Kepco 50-8M and homebuilt) were used in lieu of the regular gradient amplifiers, which were too noisy at this frequency.  The homebuilt transmit and receive coils are described in Ref [3], along with the in-house methods of HP $^3$He gas production, pulse sequence management, and data collection.  Measurements were performed using typical quantities of 50 standard cm$^3$ of ~ 30% polarized $^3$He diluted to 1 litre with $N_2$; the mixture was carried to the imager in a Tedlar plastic bag (Jensen Inert Products, **Coral Springs, Fl, USA**).

Measurements were first performed in the bag to characterize $B_0$ field homogeneity and tipping pulse quality.  Loading the transmit coil with a human subject did not change the quality factor (*Q = 12)* and, therefore, did not appreciably degrade the pulses. In vivo measurements were performed, with informed consent on 3 healthy subjects, concordant with approved protocol (CNRS 01 048).    While lying supine in the imager, the subject exhaled normally, inhaled the gas directly from the bag, then inhaled air to fill the lung and held his breath (typically 5 s or less) for signal acquisition.

Imaging was performed using a single-shot, spin-echo RARE sequence [6], while transverse relaxation times were measured using a CPMG sequence [3,6]. For all sequences, an initial $90^o_x$ RF pulse (0.5 ms) was followed by a train of $180^o_y$ RF pulses (1.0 ms) at regular interval $T_{cp}$. In a uniform field gradient $G$, the decay time $T_2'$ of the transverse magnetization in a CPMG sequence is given by

$$\frac{1}{T_2'} = \frac{1}{T_2} + \frac{D(\gamma k G T_{cp})^2}{12} \qquad (1)$$

where $T_2$ is the intrinsic relaxation time of the system (with $G = 0$) and $D$ is the coefficient of diffusion [7]. In vivo, an apparent diffusion coefficient (ADC) replaces $D$. The factor $k$ corrects for the time variation of the gradients, which were switched off during RF pulsing to obtain spatially non-selective tipping angles.

## Results

Preliminary CPMG measurements made on the bag in zero applied gradient showed the $B_0$ field to be of good homogeneity and the RF tipping pulses to be extremely accurate. For a given $T_{cp}$, values of $T_2'$ measured in the bag were an order of magnitude longer here than in our homebuilt imager [3]; this confirmed that improved homogeneity at 3 mT (of order 700 ppm) was attainable in switching to the Sopha-Magnetech imager. At the shortest inter-echo time $T_{cp} = 4.2$ ms, thousands of spin-echoes could be re-focused, and $T_2'$ values exceeded 1000 s (of order the wall-induced longitudinal relaxation time $T_1$).. If, however, the decay time is assumed to be dominated by the effects of imperfect RF pulses (Ref [3], Appendix A) then a $T_2' \sim 1000$ s puts an upper limit of $\sim 4$ ppm for the loss in transverse magnetization per $180^o$ spin-flip, which confirms the quality and accuracy of the RF pulses.

### 1D Diffusion Mapping

Typical diffusion maps made along the $x$-, $y$- and $z$-axes (right to left, back to front and foot to head respectively) are shown in Figure 1 along with FFTs of the initial spin echoes, which give a 1D profile of the gas distribution at the begining of the sequence.

Values of ADC were extracted via a pixel-by-pixel analysis of the decays of the FFTs [3]. Regional differences could be resolved, and were most notable in the $y$- and $z$-directions. Along $z$, ADC was larger in the apices of the lung. Along $y$, ADC was uniform except for a sharp increase at both extremities, which differs from the anterior to posterior ADC decrease measured at 1.5T [8]. This may be due to the difference in measurement method and probed length scale (0.6 mm in this experiment).

<div style="border:1px solid">

**Figure**

</div>

## 2D Projection Imaging

A coronal projection image, along with its 1D projections on the x- and z-axes, is shown in Figure 2. A single-shot, spin-echo sequence (RARE) was used to make maximum efficiency of the polarization [6]. The 3 mT image is of modest quality: similar to the pioneering 1.5 T images made seven years ago [9], yet a marked improvement from the first very low field image at 15 mT which suffered from poorly shielded external noise [4]. Still, we anticipate that image quality will quickly improve with better control over other noise sources (via improved low frequency filtering of lines into Faraday cage) and the use of larger quantities of $^3$He (300 cm$^3$ with 40-50% polarization, as routinely used at 1.5T).

<div style="border:1px solid">

**Figure**

</div>

## Measuring O$_2$ Partial Pressure with Fast Repetition CPMG Sequence in Zero Applied Gradient

Taking advantage of the negligible power deposition at very low RF frequency, fast repetition spin-echo sequences were used to monitor the MR signal decay with $G = 0$ in a single subject. At very low magnetic field, where susceptibility gradients play no role, the CPMG relaxation time of $^3$He in the lung is constant at small $T_{cp}$ and is only limited by the oxygen dependent magnetization lifetime $T_1$ [3]. This was observed here for $T_{cp} <$ 50 ms, where $T_2{'}$ values levelled off at ~ 25 s.  The resulting long decay time can be seen in Figure 3 where half of the initial magnetization still remains after 19s, the point at which the subject chose to exhale.

$1/T_1$ of $^3$He is known to be proportional to the oxygen partial pressure $P_{O2}$ by a constant $\xi$ = 3.8 x $10^{-4}$ $s^{-1}$/mbar at 37 $^o$C [10].   Signal decay was fit assuming a constant uptake of $O_2$ into the blood (ie. $P_{O2} = P_o - Rt$) as observed in $^3$He MRI measurements of $T_1$ at 1.5 T [11].  This nonexponential time dependence fits the data very well.  For the data of Figure 3, thenonlinear least squares fit gave an initial $O_2$ partial pressure $P_o$ of 106.7 (5) mbar and an uptake rate $R$ of 1.01 (3) mbar/s.  The high degree of certainty in the fit parameters stems from the large number of data points (~ 3200 echoes) despite a relatively low SNR (~50).  To accommodate the fast RF pulse rate here, pre-amplifier gain had to be reduced at the expense of SNR. When this technical difficulty is overcome, and if more $^3$He gas is used, SNR values up to 1000 will become routine, thereby allowing $P_o$ and $R$ to be reliably extracted with a few seconds of data only (this is shown in the inset panels of Figure 3 for simulated data).

## Figure

## Conclusion

In this article, we have presented 1D gas diffusion mapping as well as 2D imaging  of the human lung using hyperpolarized $^3$He at 3 mT.  The results of the diffusion mapping are sufficiently accurate as to distinguish regional differences in ADC.  The technical difficulties limiting image quality in this study should be easy to overcome and improvements  realized quickly.

We have also presented measurements of global $O_2$ partial pressure and uptake rate made using a fast repetition CPMG sequence in zero applied gradient.   With a readily achievable increase in SNR, these measurements will only require a few seconds to perform and use up only a fraction of the $^3$He magnetization. This opens up the very exciting possibility of performing a regular imaging sequence (regional gas diffusion or distribution map) as well as a high-accuracy global $O_2$ determination ($P_o$ and $R$) all within a single breath hold.This approach will further contribute to reduce examination time and to make for a most efficient use of $^3$He, which is a non-renewable resource. To extend this technique so as to obtain regional $O_2$ information as well, we envision the use of small local coils or SQUID detectors arrayed across the chest.

## Acknowledgements

The authors kindly acknowledge V. Senaj and G. Guillot for assisting in the preparation of the experiment and L. Darrasse and E. Durand for assisting in the preparation of the manuscript.

Figure 1. 1-D mapping of ADC value (symbols) and [3]He density (lines, arb. units) for two supine subjects (open circles and dotted lines, closed circles and solid lines). Diffusion weighting parameters: G = 900 μT/m, $T_{ep}$ = 11 - 15 ms. Top panel: projection right to left. Middle panel: projection back to front. Bottom panel: projection foot to head.

Figure 2. Coronal 2-D projection image of third subject (SNR ~ 100, resolution 6 × 18 $mm^2$, acquisition time 0.8 s). Image projections along x- and z-axes show [3]He distribution profiles similar to the 1-D images in Figure 1.

Figure 3. Nonexponential decay of fast repetition CPMG sequence ($T_{ep}$ = 6ms) in zero applied gradient. Main panel: Normalized echo amplitude versus time. The solid line is a fit to the data of the form $\exp(-\xi P_O t + \xi R t^2)$. Inset panels: Calculated relative uncertainties of parameters $\xi P_o$ and $\xi R$ as a function of the time interval used in the fit of simulated data with SNR = 50 (solid lines), 250 (dashed lines), 1000 (dotted lines).

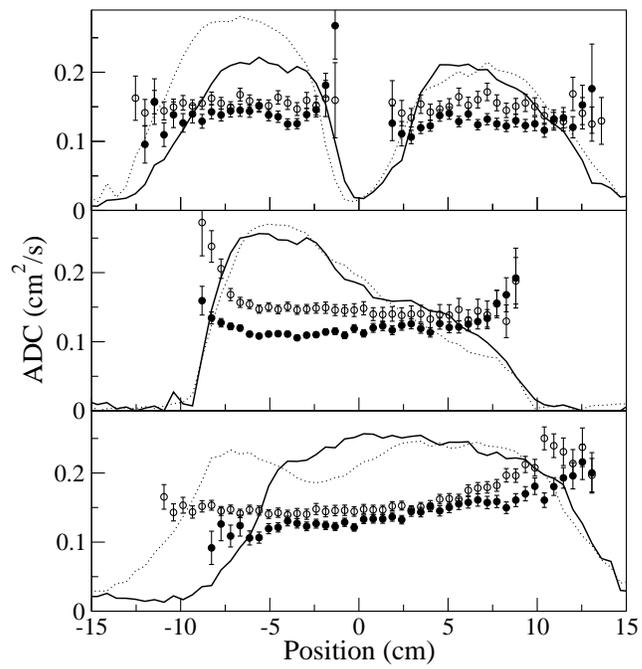

Figure 1:

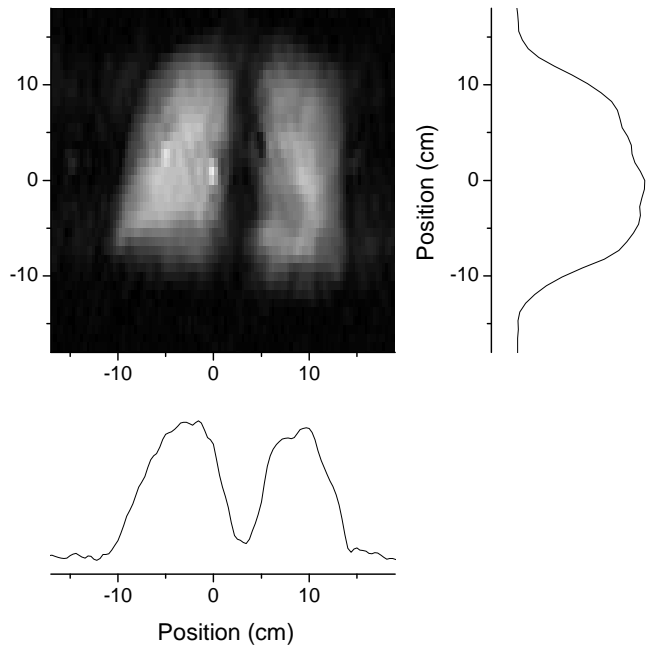

Figure 2:

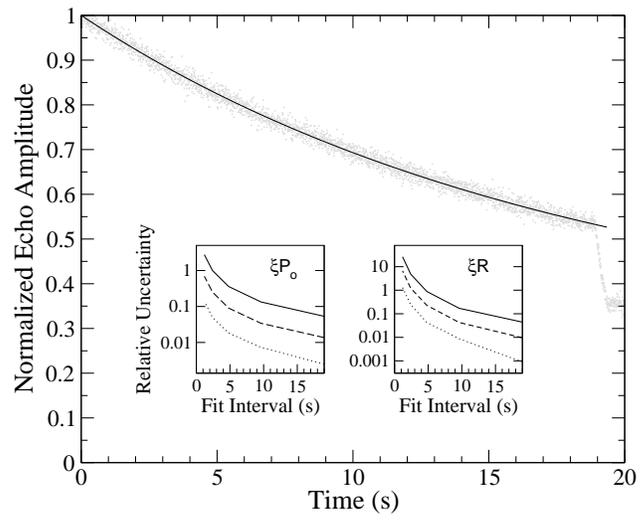

Figure 3: